\documentclass[12pt]{article}
\usepackage{aas_macros}
\usepackage{times}
\usepackage{geometry}
\usepackage[utf8]{inputenc}
\usepackage{enumitem,amssymb}
\usepackage{ragged2e}
\usepackage{graphicx}
\usepackage{fancyhdr}
\usepackage{hyperref}
\usepackage{tabularx}
\usepackage[
    natbib=true,
    style=numeric,
    sorting=none
]{biblatex}
\usepackage{color, colortbl}
\newlist{thematic}{itemize}{8}
\setlist[thematic]{label=$\square$}
\usepackage{pifont}

\newcommand\farcs{\mbox{$.\!\!^{\prime\prime}$}}%

\begin{document}


\raggedright
\huge{
A Roman Coronagraph Spectroscopic Mode Demonstration}
\linebreak
\large

Thayne Currie$^{1,2}$,Brianna Lacy$^{3}$,
Yiting Li$^{4}$, 
Mona El Morsy$^{1}$, Danielle Bovie$^{1}$, Kellen Lawson$^{5}$,  Masayuki Kuzuhara$^{6}$, Naoshi Murakami$^{6}$

1 Department of Physics and Astronomy, University of Texas-San Antonio, San Antonio, TX, USA \\
2 Subaru Telescope, National Astronomical Observatory of Japan, 
Hilo, HI, USA\\
3 Department of Astronomy \& Astrophysics, University of California-Santa Cruz, Santa Cruz, CA, USA\\
4 Department of Astronomy, University of Michigan, Ann Arbor, MI, USA\\
5 NASA Goddard Spaceflight Center, Greenbelt, MD, USA\\
6 National Astronomical Observatory of Japan, 2-21-1 Osawa, Mitaka, Tokyo 181-8588, Japan

\justify{
\textbf{Summary:} We propose 730 nm high-contrast spectroscopic observations of selected self-luminous directly-imaged planets as a key test of the Roman Coronagraph's planet characterization capabilities.  The planet sample draws from ground-based IR discoveries with the NASA headquarters-supported Subaru/OASIS survey -- HIP 99770 b and HIP 54515 b --  and ``emblematic" planets $\beta$ Pic b and HR 8799 e.  All of these planets are likely unsuitable for achieving the coronagraph's core TTR5 goal at 575 nm but are detectable at longer wavelength passbands.
 Their predicted contrasts at 730 $\mu m$ cover two orders of magnitude range; all companions reside within the dark hole region enabled by the shaped-pupil coronagraph at 730 nm.  These observations will help to fulfill multiple Coronagraph Objectives, providing a first assessment of the wavelength dependence of speckle noise and the ability to extract accurate atmospheric information in the face of this noise.   Additionally, they will provide a first experiment at extracting optical planet spectra in the face of signal contamination from a debris disk: prefiguring challenges that the \textit{Habitable Worlds Observatory} may encounter with imaging Earths in exozodi-contaminated systems.

\pagebreak
\noindent \textbf{Type of observation:} \\$\boxtimes$   Technology Demonstration\\
$\boxtimes$ Scientific Exploration\\\\

\noindent \textbf{Scientific / Technical Keywords:}  
companion (substellar), companion (exoplanet), self-luminous, high contrast performance\\

\noindent \textbf{Required Detection Limit:} \\ 
\begin{tabular}{| c | c | c | c | c |}
\hline
$\geq$10$^{-5}$ & 10$^{-6}$ & 10$^{-7}$ & 10$^{-8}$ & 10$^{-9}$ \\ \hline
 & x & x & x & \\ \hline
\end{tabular}

\vspace{0.5cm}
\textbf{Roman Coronagraph Observing Mode(s):} 

\begin{tabular}{| c | c | c | c | c |}
\hline
Band &   Mode & Mask Type & Coverage & Support \\ \hline \hline
3, 730 nm &  Slit + R$\sim$50  & Shaped Pupil & 2x65$^{\circ}$ & Best Effort \\ 
 &    Prism Spectroscopy &  &  & \\  \hline
\end{tabular}
}



\justify{
\begin{center}
\begin{tabular}{| c | c | c | c | c |}
\hline
Name &  host star & detection & separation (") & description of target \\
  & V mag. & limit & (or extent)  & \\ \hline \hline
  HIP 99770 b & 4.9 & 3$\times$10$^{-8}$ & 0.35 & self-luminous exoplanet\\ 
  $\beta$ Pic b & 3.9 & 5$\times$10$^{-7}$ & 0.35 & self-luminous exoplanet\\
  HR 8799 e & 5.9 & 7$\times$10$^{-8}$ & 0.41 & self-luminous exoplanet\\
  HIP 54515 b & 6.8 & 3$\times$10$^{-6}$ & 0.23 & self-luminous exoplanet\\
  \hline
\end{tabular}
The contrasts listed are median estimates. See Figures for the full range of predicted values.
\end{center}

}

\pagebreak


\begin{figure}[]
\centering
     \includegraphics[width=0.4\textwidth,clip]{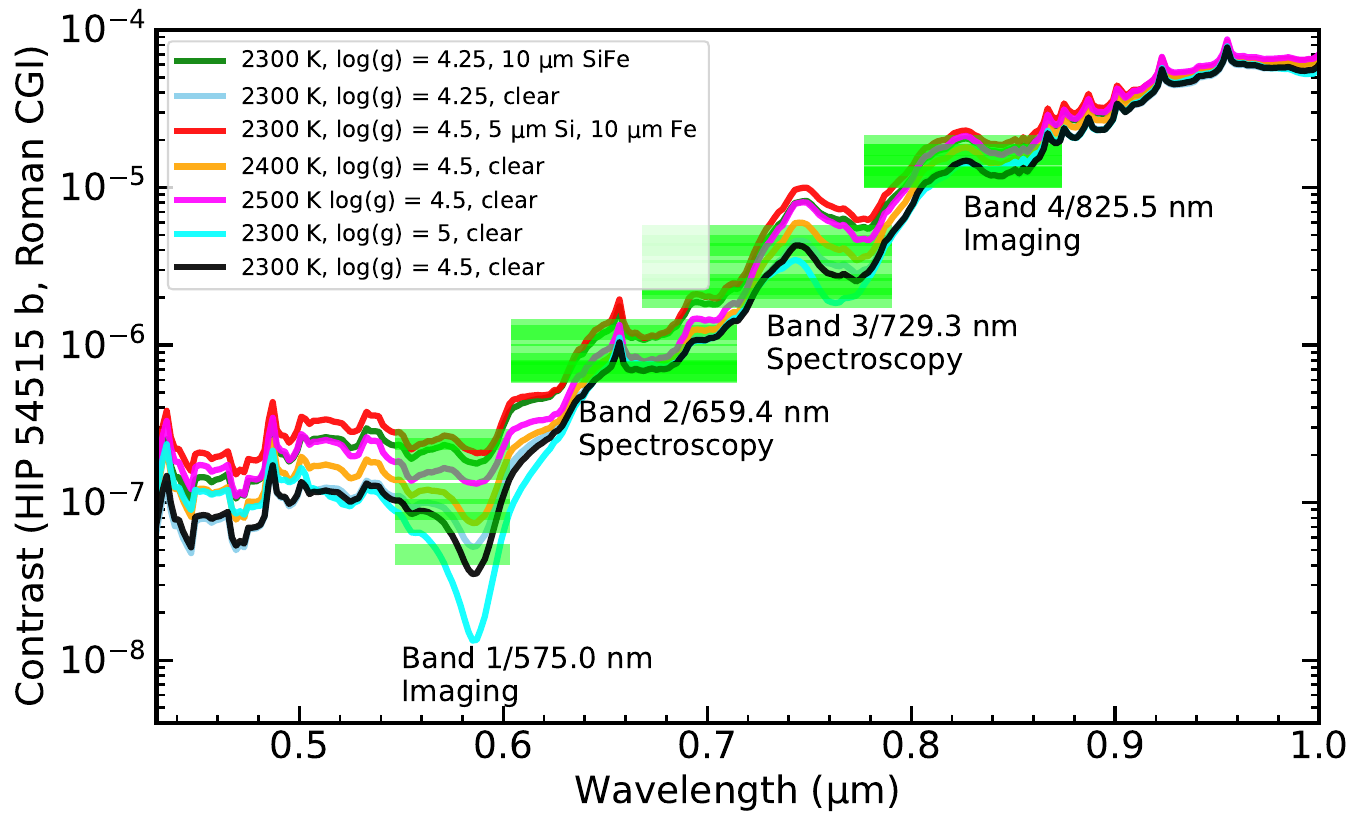}
     \includegraphics[width=0.4\textwidth,clip]{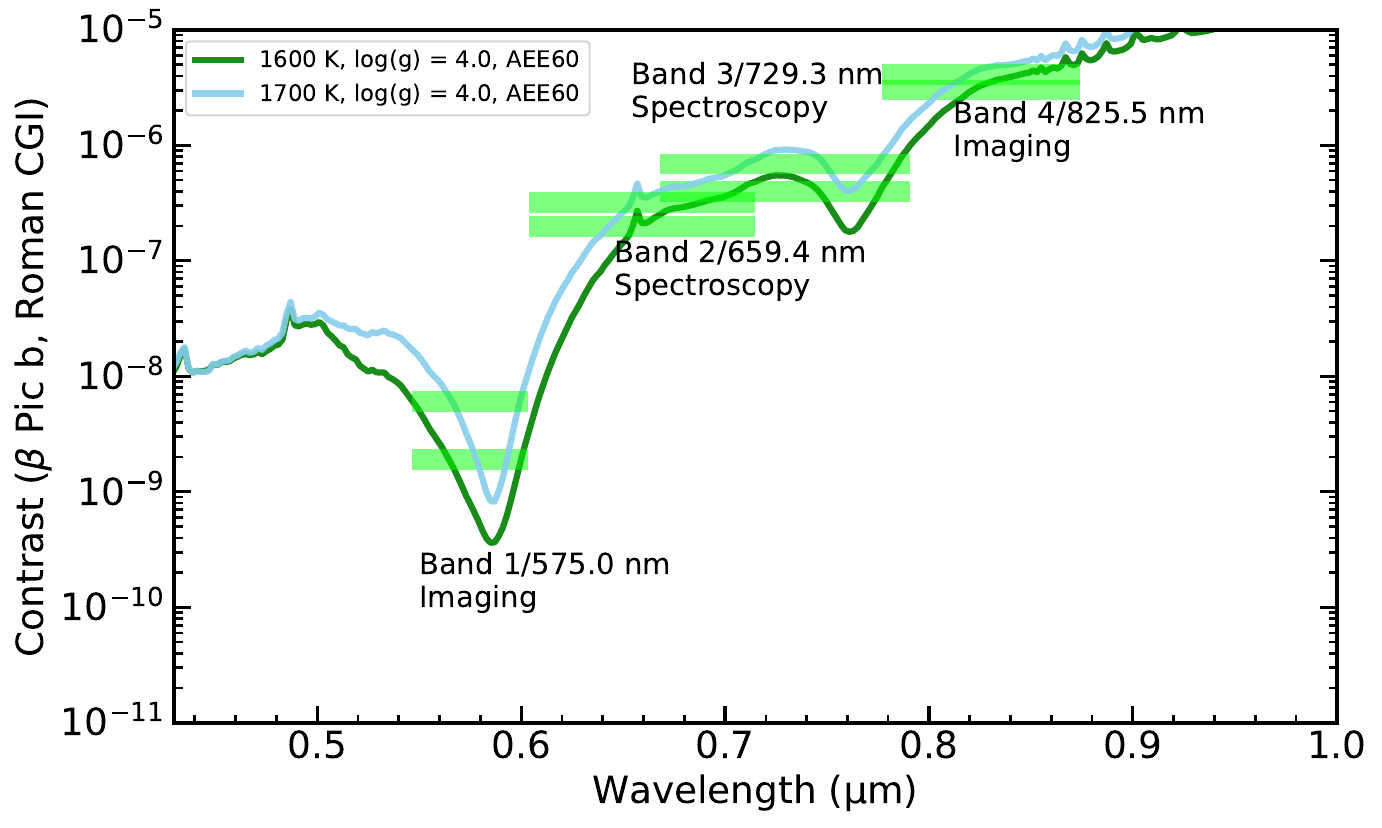}
    \includegraphics[width=0.4\textwidth,clip]{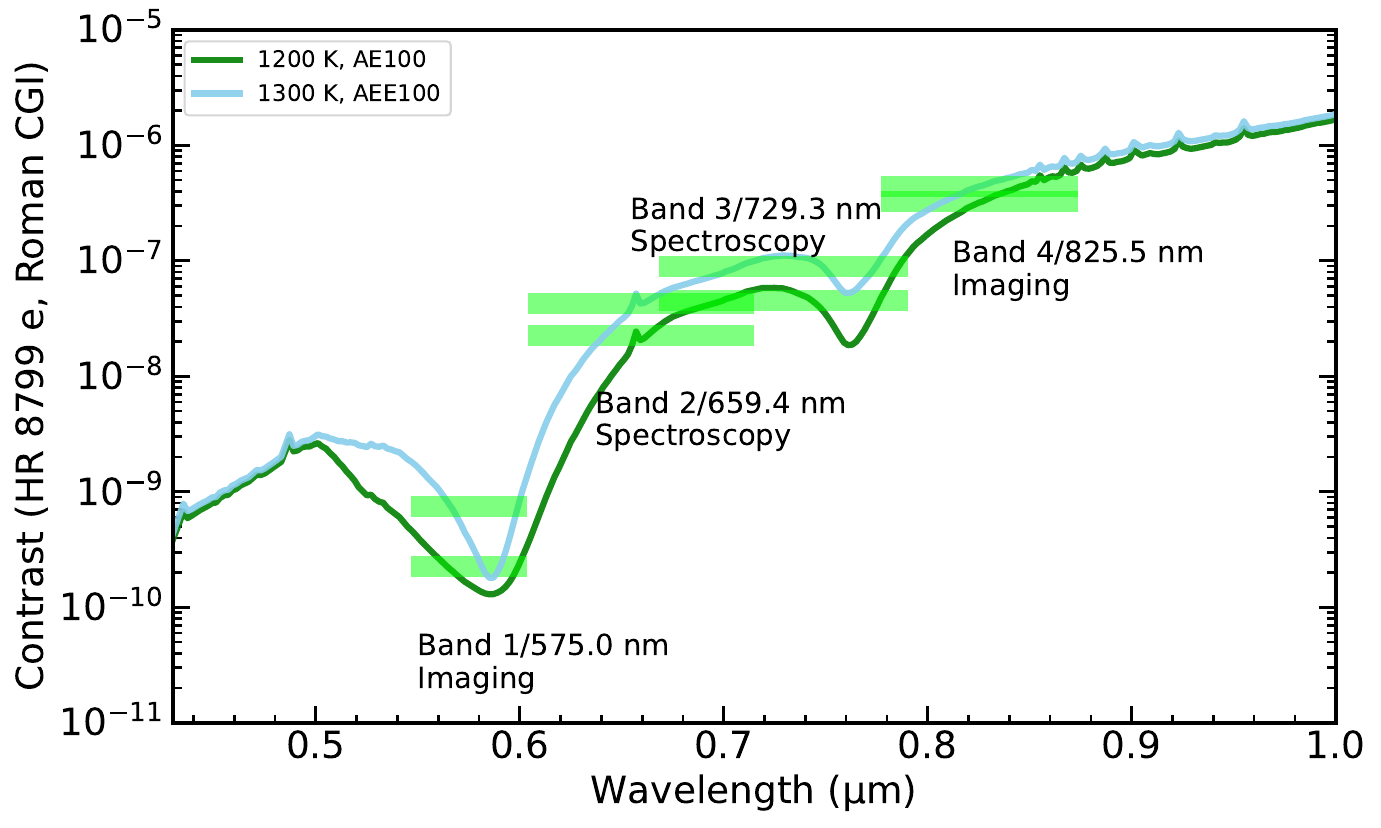}
     \includegraphics[width=0.4\textwidth,clip]{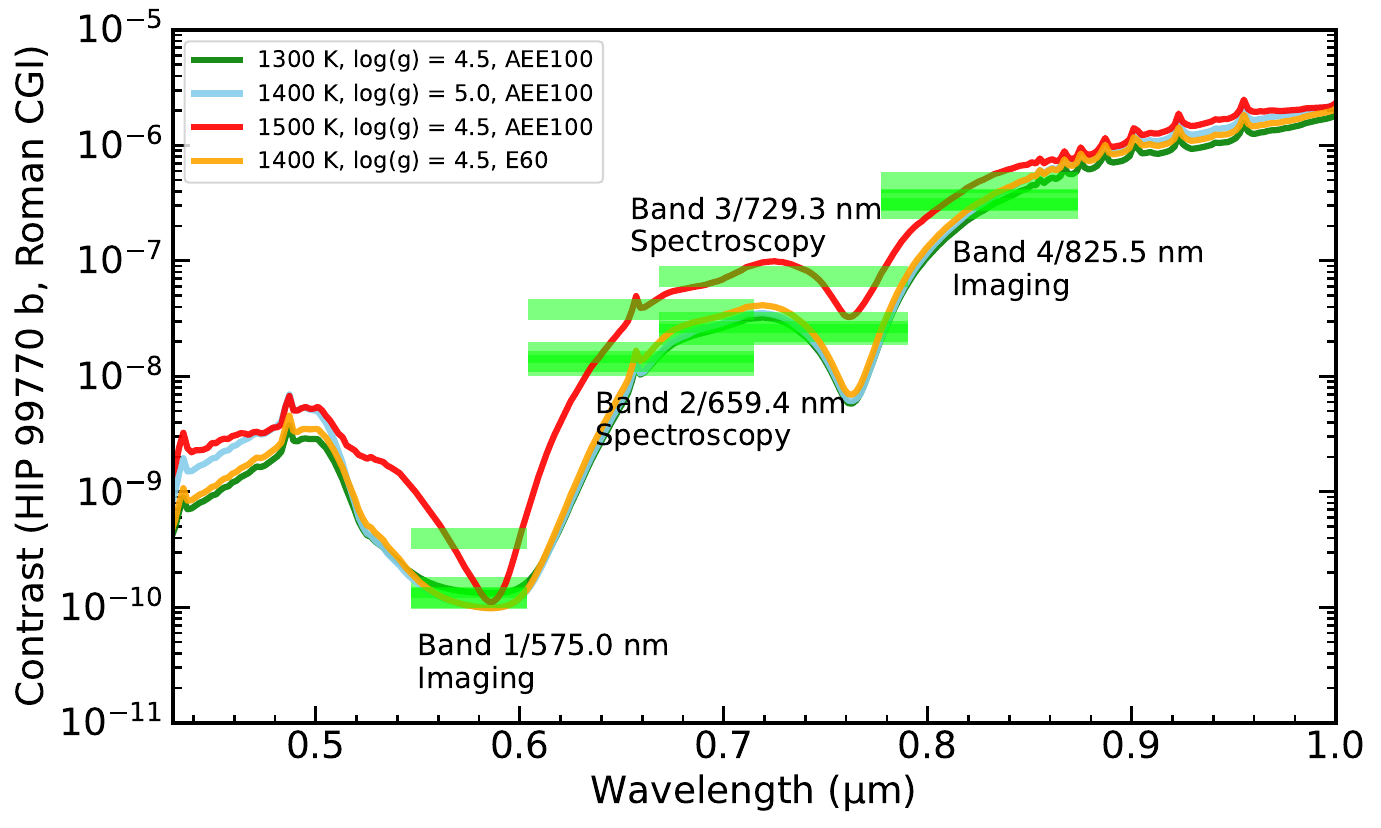}
    \vspace{-0.25in}
    \caption{ Contrast predictions for $\beta$ Pic b, HIP 54515 b, HR 8799 e, and HIP 99770 b from updates to the \citeauthor{LacyBurrows2020} atmosphere models described in \citet{Currie2023a}.   The temperatures and gravities for $\beta$ Pic b, HIP 54515 b, and HIP 99770 b draw from published atmospheric modeling of near-IR spectra and photometry \citep{Chilcote2017,Currie2025,Bovie2025}.   The assumed temperature of 1200 $K$ for HR 8799 e matches the planet's SED, where the 1300 $K$ model fares slightly worse.  $\beta$ Pic b is likely undetectable at 575 nm due to its steep contrast and confusion with the system's bright debris disk.  \textbf{However}, it should be detectable at 730 nm, where it is a factor of 30--100 times brighter relative to the star and disk: this prediction is consistent with a recent MagAO-X detection of $\beta$ Pic b at 762nm: see \url{https://nexsci.caltech.edu/workshop/2025/posters/Poster_EdenMcEwen_176.pdf}.  
    }
\end{figure}

\justify{
\textbf{\Large Anticipated Technology / Science Objectives:}
Beyond TTR5, the Roman Coronagraph instrument (CGI) is tasked with fulfilling five separate ``Objectives" (2.2.1--2.2.5).    In particular, Objective 2.2.5 requires characterizing ``\textit{photometry, \textbf{spectroscopy}, and astrometry"} of at least one companion.

The CGI spectroscopic mode at 730 nm is a powerful science enabler for any follow on CGI extended science program and prototypes future transformative science programs with the \textit{Habitable Worlds Observatory}.  For reflected-light planets, even noisy (S/N $\sim$ 5) low-resolution spectra can reveal molecules (e.g. methane) and begin to constrain their chemical abundances \citep{Burrows2014,Marley2014}.   Warm self-luminous jovian planets also likely show strong red-optical features: e.g. the potassium doublet at 0.77 $\mu m$, which is predicted to be strongly dependent on metallicity and the presence of clouds \citep{LacyBurrows2020} found to be ubiquitous from near-infrared (IR) observations \citep{Currie2011}.  The contrasts that CGI might reach in the 730 nm bandpass ($\sim$10$^{-8}$--10$^{-9}$) at 3--9 $\lambda$/D are 100-1000 times deeper than any achievable from current ground-based telescopes and within a factor of 10--100 needed to detect an Earth twin around a Sun-like star.  Thus, CGI presents a new and hitherto unexplored contrast regime for detecting planets, extracting their spectra, and deriving their atmospheric parameters.

We propose a Roman CGI ``spectroscopic mode demonstration" with the goals of 1) determining CGI's contrast and detection capabilities with long-slit spectroscopy, 2) quantifying the characteristic speckle length as a function of wavelength in different conditions (angular separation, contrast, star brightness, etc.),  3) assessing our ability to extract accurate atmosphere information in the face of this noise and in the face of contamination from a bright debris disk.

Our targets consist of directly-imaged planets discovered from ground-based extreme AO data with well-constrained atmospheric properties: HIP 99770 b, HR 8799 e, $\beta$ Pic b, and HIP 54515 b \citep{Currie2023a,Marois2010,Lagrange2010,Currie2025}\footnote{One recent submitted -- not accepted -- work claims that the transition from planets to brown dwarfs occurs at $\sim$ 10 $M_{\rm Jup}$ (Hsu et al. 2026, submitted).  Taken literally, this might imply that some of our targets are brown dwarfs.  However, recent, extensive peer-reviewed studies not considered in above work, tracing the demographics of companions through mass functions or metallicities, contradict this proposed distinction.  These studies find a transition between planets and brown dwarfs at much higher masses, typically $\approx$ 25 $M_{\rm Jup}$, and mass ratios of $q$ $\sim$ 2--3\% \citep{Currie2023a,Currie2025,Feng2022,Meyer2025,Giacalone2026}.  Additionally, the rotational velocity-sampling of companions at the 5--10 $M_{\rm Jup}$ range is sparse and does not include a substantial number of free floating objects as a control sample.  Some of these, e.g. PSO J318-22, appear to be rapid rotators that may also provide rebutting evidence \citep{Allers2016,Biller2018}.  The rotational evolution of substellar objects may be mass sensitive but it does not distinguish between planets and brown dwarfs at this time.}.    These planets include long-standing ``emblematic" planets \citep{Marois2010,Lagrange2010,Currie2023b} that happen to be amenable to Roman high-contrast imaging at 730 nm and those discovered from the Observations of Accelerators with SCExAO Imaging Survey (OASIS) (PI T. Currie; \citealt{ElMorsy2024}) funded by NASA Headquarters as \textit{Strategic Mission Support} to identify Roman Coronagraph targets for the technology demonstration phase.

A decade of astrometric monitoring from direct imaging data (HR 8799 e, $\beta$ Pic b) or strong or orbit constraints from jointly modeling direct imaging and astrometry (HIP 99770 b, HIP 54515 b) constrains the companions' positions during the Roman CGI technology demonstration phase.  The latest orbital fits to HR 8799 e and $\beta$ Pic b predict that the planets will be at $\rho$ $\sim$ 0\farcs{}35 in November 2027 and January 2028, respectively\footnote{See \url{http://whereistheplanet.com/}}.   HIP 99770 b and HIP 54515 b will lie at $\rho$ $\sim$ 0\farcs{}35 and 0\farcs{}23, respectively \citep{Bovie2025,Currie2025}.

Figure 1 displays the full range of predicted contrasts for each target in the Roman CGI passbands.   Even absent contamination from debris disk emission, HR 8799 e has a predicted 575 nm contrast below 10$^{-9}$, rendering it likely undetectable with CGI.   Some recent work posits that $\beta$ Pic b may be a workable technology demonstration target for achieving TTR5 \citep{Hom2025}.  However, its steep predicted contrast ($\approx$ 2--6$\times$10$^{-9}$) combined with contamination from the system's bright debris disk likely renders the planet undetectable at 575 nm even in the most optimistic CGI performance scenario.  HIP 99770 b has a temperature comparable to HR 8799 e but is less cloudy, which deepens its sodium absorption feature \citep{LacyBurrows2020}.  As a result, its 575 nm contrast is even steeper than HR 8799 e's.   HIP 54515 b's predicted 575 nm contrast is $\approx$10$^{-7}$. But in addition to the companion's small angular separation, the primary has an optical brightness of V = 6.8.   Other targets -- e.g. HIP 71618 B -- are better suited for achieving TTR5.

While not well suited for demonstrating the Coronagraph's TTR5 requirement, our sample's predicted 730 nm contrasts span a range of feasible CGI performances at 730 nm (10$^{-6}$ -- 10$^{-8}$) The primaries vary in brightness between V $\approx$ 4 and V $\approx$ 7.  Thus, targeting these stars with CGI tests our ability to perform spectral extraction and atmospheric retrieval over a range of target properties.  All have dynamical mass measurements and constrained ages: analysis of their CGI spectra will then help inform how the atmospheres of planets evolve at different ages.
The easiest to detect targets (HIP 54515 b, $\beta$ Pic b) are suitable as ``first-look"/commissioning observations.   The others require contrasts below the TTR5 level of $\sim$10$^{-7}$ and are thus better suited for the observation phase soon after TTR5 is achieved on another star.  

As of now, the only demonstrably suitable target for TTR5 demonstration is HIP 71618 \citep{ElMorsy2025}.  Scheduling of spectroscopic mode demonstration observations is then likely contingent upon the observing schedule for HIP 71618.

}


\textbf{\Large Observing Description:}
The program goal requires Band 3/730 nm long-slit spectroscopy of planets and brown dwarfs (listed above) with the shaped pupil coronagraph.  The observations should follow the ``standard typical observing sequence" described on page 36 of the most recent CGI white paper slide deck that enables both angular and reference star differential imaging (ADI, RDI)\footnote{Coronagraph$\_$CPP$\_$WPoverview2025$\_$8July2025}.  Each "visit" consists of dark hole digging on a bright nearby PSF reference, PSF reference observations, two sets of +/- telescope rolls ($\Delta$$\theta$= +/- 15$^{o}$) on the target star, followed by a second set of PSF reference observation.  

The program will demonstrate spectral extraction of a planet in a range of contamination levels from exozodaical light.     HIP 54515 lacks evidence for exozodi emission or a debris disk whose scattered light could contaminate their companions' spectra.  While HIP 99770 has a cold debris belt, the system's nearly face-on viewing geometry makes this dust population an unlikely contaminant.  HR 8799 is surrounded by a warm debris population imaged with JWST in the thermal IR and potentially in near-IR polarimetry \citep{Boccaletti2024,Engler2025}, but there are no such optical detections thus far: weak contamination is a small possibility. $\beta$ Pic b is likely rendered undetectable at 575 nm due to the system's bright cold debris disk.  While it is detectable at 730 nm (see Fig. 1 caption), the debris disk will very likely contaminate the signal.  All targets have suitable (candidate) PSF reference stars with small $\Delta$ pitch angles.  As an example, we display the $\Delta$ pitch angle vs time (in 2027) for HIP 54515, which has HIP 54872 and HIP 57632 as possible references (Figure 2).

\begin{figure}[!h]
\centering
 \includegraphics[width=0.275\textwidth,clip]{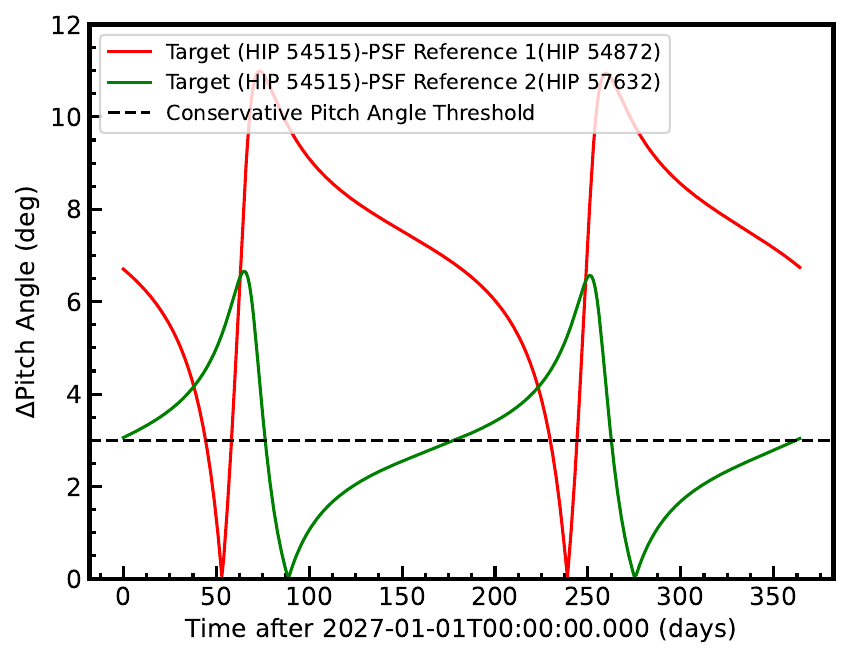}
      \includegraphics[width=0.355\textwidth,clip]{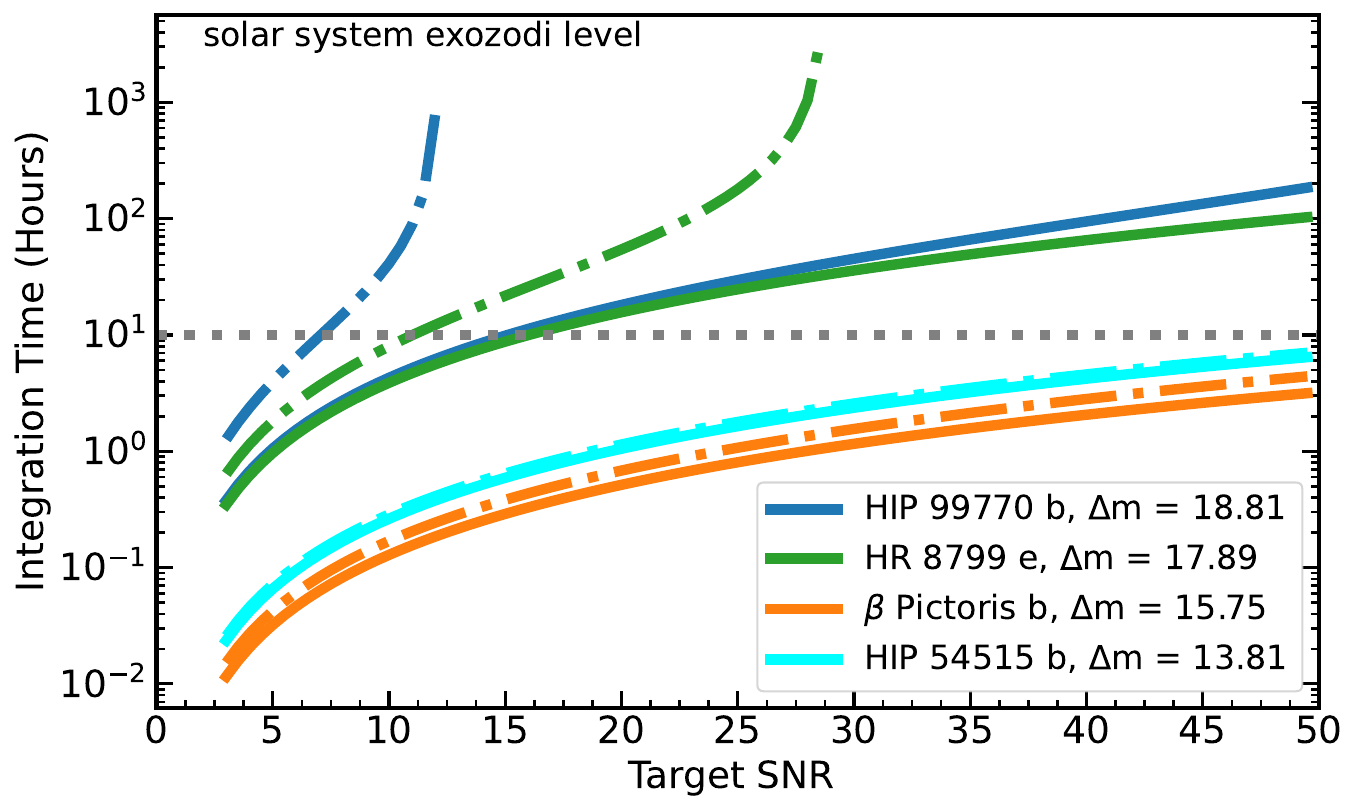}
      \includegraphics[width=0.355\textwidth,clip]{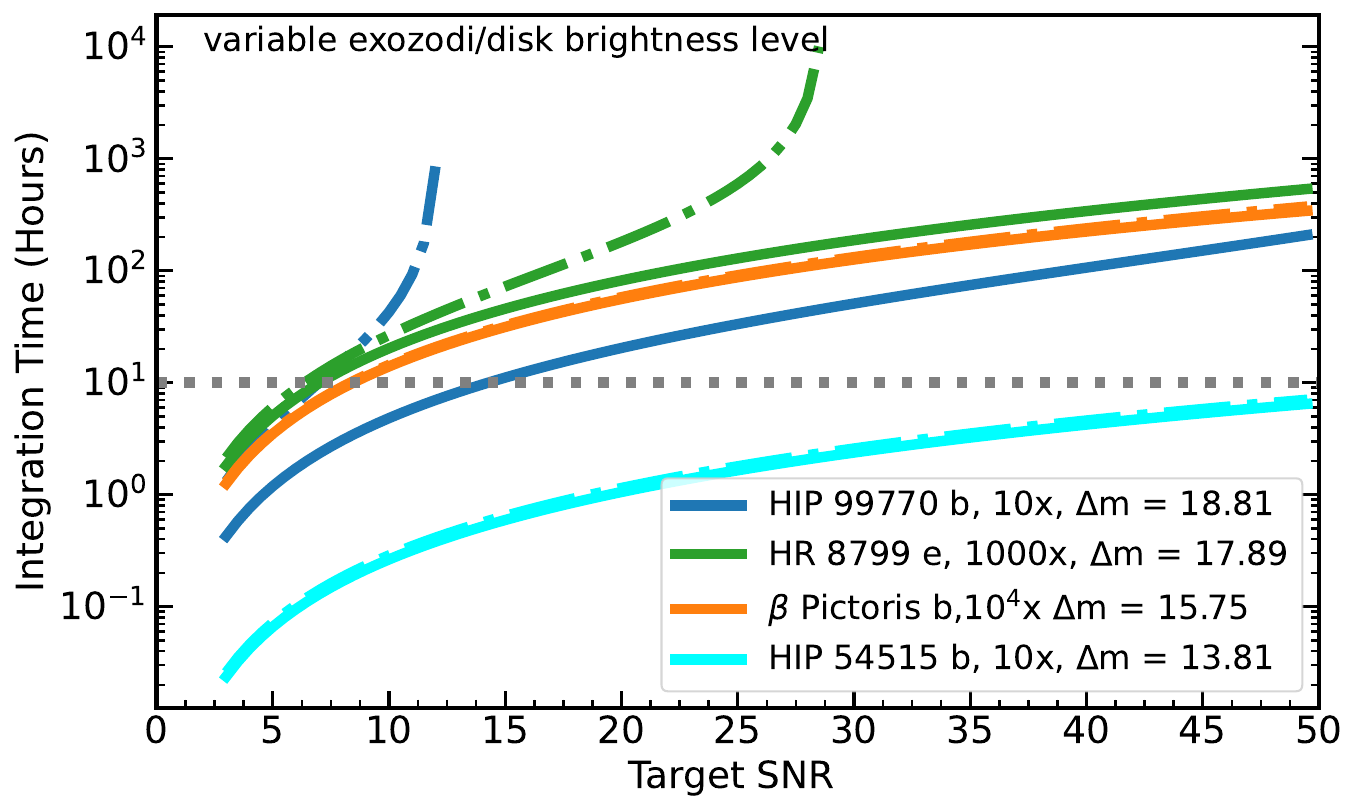}
    \vspace{-0.35in}
    \caption{(left) Relative Pitch Angle between HIP 54515 and two potential reference stars during the calendar year of 2027.  Combined with keepout maps (not shown), the second PSF reference (HIP 57632) would allow observations with a $\Delta$ Pitch Angle less than 3$^{o}$.  (middle, right) SNR vs. integration time for our targets with solar levels of exozodi light and variable ones assuming conservative (dash-dot) or optimistic (solid) coronagraph performances.  The latter are either set at 10 times solar system levels or are crudely matched to the system's debris disk brightness.
    }
\end{figure}
\textbf{Estimate of Time Needed}: We use the Roman Coronagraphic Instrument Exposure Time Calculator (Corgi-ETC) to estimate our achieved SNR vs. integration time.   We set the contrast equal to the mean predicted contrast for each companion as listed above (e.g. 3$\times$10$^{-6}$ for HIP 54515 b and 3$\times$10$^{-8}$ for HIP 99770 b) and calculate predicted performance for both optimistic and conservative scenarios for CGI.  We assess the impact of exozodi or debris disk contamination by 1) either increasing the exozodi level to 10 times solar system values (HIP 99770 b, HIP 54515 b) or raising it to yield a surface brightness comparable to the HR 8799 warm dust population and $\beta$ Pic's debris disk and 2) setting all values to one exo-zodi.   We plan to observe each target for $\sim$10 hours clock time in order to provide a uniform assessment of performance.    In 10 hours of integration time (17 hours clock time), our predicted SNRs for conservative to optimistic cases are then 7--15 (7-15), 7--8 (11-15), 7--8 ($>$50), and $>$50 ($>$50) for HIP 99770 b, HR 8799 e, $\beta$ Pic b, and HIP 54515 b in the limit of strong (weak) exozodi/debris disk contamination (Fig. 2).




\pagebreak

\printbibliography

\end{document}